\documentclass[fleqn,10pt]{wlscirep}
\usepackage[utf8]{inputenc}
\usepackage[T1]{fontenc}
\usepackage[acronym, shortcuts]{glossaries}
\usepackage{siunitx}
\usepackage{todonotes}
\usepackage[version=4]{mhchem}
\newacronym{CPT}{CPT}{coherent population trapping}
\newacronym{MEMS}{MEMS}{micro-electro-mechanical systems}
\newacronym{FWHM}{FWHM}{full width at half maximum}
\newacronym{DRIE}{DRIE}{deep reactive ion etching}

\title{Laser-actuated hermetic seals for integrated atomic devices}

\author[1]{Vincent Maurice}
\author[2]{Clément Carlé}
\author[2]{Shervin Keshavarzi}
\author[1]{Ravinder Chutani}
\author[2]{Samuel Queste}
\author[2]{Ludovic Gauthier-Manuel}
\author[2]{Jean-Marc Cote}
\author[2]{Rémy Vicarini}
\author[2]{Moustafa Abdel Hafiz}
\author[2]{Rodolphe Boudot}
\author[2,*]{Nicolas Passilly}

\affil[1]{IEMN - Institut d'Electronique de Micro\'{e}lectronique et de Nanotechnologie,  UMR8520 CNRS, Universit\'{e} Lille, Centrale Lille, Lille, F-59000, France}
\affil[2]{FEMTO-ST Institute, UMR6174 CNRS, Universit\'{e} Bourgogne Franche-Comt\'{e}, Besançon, F-25030, France\\}

\affil[*]{nicolas.passilly@femto-st.fr}

\begin{abstract}

Atomic devices such as atomic clocks and optically-pumped magnetometers rely on the interrogation of atoms contained in a cell whose inner content has to meet high standards of purity and accuracy. Glass-blowing techniques and craftsmanship have evolved over many decades to achieve such standards in macroscopic vapor cells.
With the emergence of chip-scale atomic devices, the need for miniaturization and mass fabrication has led to the adoption of microfabrication techniques to make millimeter-scale vapor cells. However, many shortcomings remain and no process has been able to match the quality and versatility of glass-blown cells.
Here, we introduce a novel approach to structure, fill and seal microfabricated vapor cells inspired from the century-old approach of glass-blowing. In particular we demonstrate opening and closing single-use zero-leak microfabricated valves, actuated exclusively by laser, and operating in the same way as the “make-seals” and “break-seals” found in the filling apparatus of traditional cells. Such structures are employed to fill cesium vapor cells at the wafer-level. 
The make-seal structure consists of a glass membrane that can be locally heated and deflected to seal a microchannel. The break-seal is obtained by breaching a silicon wall between cavities. This new approach allows adapting processes previously restricted to glass-blown cells. It can also be extended to vacuum microelectronics and vacuum-packaging of \ac{MEMS} devices.

\end{abstract}
\begin{document}
\flushbottom
\maketitle
\thispagestyle{empty}

\section*{Introduction}

The inherently invariable structure of the atoms has allowed the development of instruments featuring tremendous performances.
Because their structure relates to physical quantities through constants in the most direct way, probing the states of atoms with lasers provide an effective way of measuring such quantities with high accuracy and sensitivity.   
Atom-based frequency references with fractional uncertainties down to the $10^{-20}$ level~\cite{bothwellResolvingGravitationalRedshift2022}, optically pumped magnetometers able to measure neuronal activity~\cite{botoMovingMagnetoencephalographyRealworld2018} and high-sensitivity terahertz electric field imaging~\cite{wadeRealtimeNearfieldTerahertz2017} are compelling feats of atomic devices.
Atoms, both in vapor-form or in laser-cooled ensembles, also offer an appealing platform for quantum information processing and sensing as it allows operating near room temperature~\cite{degenQuantumSensing2017}.

While state-of-the-art atom-based instruments keep breaking records, they are usually complex, cumbersome and expensive.
Instead, numerous applications require distributed or remote measurements using sensors that are accessible, arrayable and easy to deploy.
In such cases, size, weight, power and cost constraints also need to be addressed and research efforts have intensified to this end.
In the early 2000s, a new approach was proposed at NIST to drastically miniaturize atomic clocks~\cite{knappe_microfabricated_2004-1}.
It consisted in using a laser diode to probe atoms confined within a vapor cell fabricated, for the first time, using microfabrication techniques.
To make this cell, a through-cavity was formed in a silicon wafer using \ac{DRIE} before joining glass wafers on both sides with anodic bonding.
The resulting device was exclusively comprised of chip-integrated components and this demonstration opened the path toward small and inexpensive atomic devices.
Later on, chip-scale atomic devices have found remarkable uses such as high-sensitivity magnetometers~\cite{shah_subpicotesla_2007}, nuclear magnetic resonance signal detection~\cite{ledbetterZerofieldRemoteDetection2008} and optical clocks~\cite{newmanArchitecturePhotonicIntegration2019}.
Several devices have already been brought to market and prospects are numerous~\cite{Kitching2018}.

To reach high performances, atoms need to be isolated in a pristine environment that avoids parasitic collisions with background molecules or against the inner walls of the container in which the atoms are confined.
Furthermore, this constraint worsens as the size of the vapor cell decreases. 
Laboratory-scale atomic devices have been using glass vapor cells and ultra-high-vacuum chambers to provide such an environment, fully benefiting from refined glass-blowing techniques, a wide variety of materials and advanced vacuum technology.
Conventional glass-blown cells have allowed anti-relaxation coatings with record performances\cite{balabas_polarized_2010}, low background gases able to sustain Rydberg states suitable for electric field sensing\cite{jingAtomicSuperheterodyneReceiver2020} or single photon sources\cite{ripkaRoomtemperatureSinglephotonSource2018}, all fabricated with a unique and versatile technique.

However, when it comes to integrated devices, the solutions available to provide such pristine environments in microfabricated structures are far more limited.
The difficulties arise in part from the high reactivity of the alkali metals often used in such devices, which limits the range of materials, processes and temperatures that can be used.
Anodic bonding has been key in the fabrication of microfabricated cells as it provides a way of joining the glass windows on both sides of an etched silicon wafer, which define the inner volume of the cell, while, at the same time, hermetically sealing it.
Although anodic bonding has become the go-to solution to seal the cells, filling them with alkali vapor remains challenging and no single solution has been able to address the needs of all atomic devices.
Instead, various filling solutions have been proposed, each with its pros and cons.

A first set of methods involves inserting pure alkali metal directly into the cell preforms before anodic bonding.
This can be done by evaporating or pipetting the metal in an anaerobic chamber or a vacuum chamber~\cite{liew_microfabricated_2004, knappe_chip-scale_2005, bopp_wafer-level_2020}.
However, anodic bonding requires elevated temperatures (250 to \SI{350}{\celsius}), which can cause the alkali metal to evaporate away from the preform or diffuse into the glass\cite{brossel_absorption_1955}.
In addition, the combination of heat and alkali vapor can damage dielectric optical coatings\cite{li_optical_2011}.

Instead of manipulating pure substances, solutions where stable compounds are decomposed into elemental alkali metal after sealing have provided valuable alternatives.
This includes the use of alkali azides (\ce{RbN_3}, \ce{CsN_3}), which can be decomposed through UV light exposure\cite{woetzel_microfabricated_2011,liew_wafer-level_2007}, and alkali chromates or molybdates dispensers in the form of solid pills\cite{douahiVapourMicrocellChip2007, hasegawa_microfabrication_2011-1, Vicarini2018} or paste\cite{maurice_microfabricated_2017}, which can be activated by laser-heating once the cell is sealed.
However, both solutions come with their own limitations: alkali azides tend to release large amounts of nitrogen, which can be undesirable, and alkali dispensers prevent non-noble gases from being inserted, due to the gettering compound contained in the pills.
In addition, dispensers remain bulky and expensive, which in turn imposes upper bounds on the footprint and cost of the cell.

In spite of its robustness and hermeticity, anodic bonding, on its own, also imposes specific restrictions, regardless of the method used to fill the cells.
First, it releases oxygen, which can react with alkali metals and decrease the lifetime of the cell.
Second, the elevated temperature makes it prone to outgassing and prevents heat-sensitive substances from being sealed or antirelaxation coatings from being used.
Alternatives based on indium or Cu-Cu thermocompression bonding have been investigated but still suffer from short lifetimes due to the reaction between indium and alkali metals in the first case and the elevated temperature still required in the latter case.\cite{straessle_microfabricated_2014, karlen_sealing_2019}

In contrast to the variety of methods that have been proposed to make microfabricated alkali vapor cells, macroscopic cells have been able to meet stringent quality and purity specifications through a unique and versatile method based on glass-blowing.
This method consists in connecting a set of empty glass cells to a manifold providing access to a source of alkali metal, gas cylinders and a vacuum pump.\cite{Singh1972}
In a typical fabrication run, the cells are first evacuated to clear the atmosphere.
Alkali metal is then migrated from the source within each cell using a torch.
Once the desired atmosphere is established (vacuum or buffer gas), the cells are "pinched-off" by melting the thin capillary that connects them to the manifold and pulling the cell away. This method relies on two key components allowing to connect or hermetically seal inner chambers at different stages within the process.
In particular, a "break-seal" apparatus, usually made of a magnetic ball thrown against a two-compartment separation membrane, is employed to free the alkali vapor from the source so that it can migrate toward the cell.
A "make-seal" apparatus allows the cell to be sealed and pinched-off after filling.
Fig.~\ref{fig:schemes}a shows a typical glass assembly used for filling cells and illustrates the sealing process.
Since the cells can be evacuated, filled and sealed after having been fully formed, this technique is prone to higher vacuum levels.
In addition, it is adequate for heat-sensitive coatings because the heat applied while sealing the cell remains localized to the connecting tube.

In this work, we propose a novel approach to bring the intrinsic benefits of glass-blown vapor cells to microfabricated cells, well-suited for miniaturization and mass-production, and demonstrate micro-integrated versions of the two key components found in conventional cell filling setups: the make-seal and the break-seal.
Both components are actuated by lasers, making batch-processing easier than their glass-blown counterparts.
We present the design and the fabrication of the structures and report on their long-lasting hermeticity through spectroscopic measurements.
As proof of principle, we show how they can be used to fill a batch of cells from a single alkali source or to fine-tune the buffer-gas content of individual cells.
This approach lets us envision adapting a wide variety of processes that could not be done at scale yet.

\section*{Results}
%   ____  _____ ____  _   _ _   _____ ____  
%  |  _ \| ____/ ___|| | | | | |_   _/ ___| 
%  | |_) |  _| \___ \| | | | |   | | \___ \ 
%  |  _ <| |___ ___) | |_| | |___| |  ___) |
%  |_| \_\_____|____/ \___/|_____|_| |____/ 

\begin{figure}[!htp]
	\centering
	\includegraphics[width=17cm]{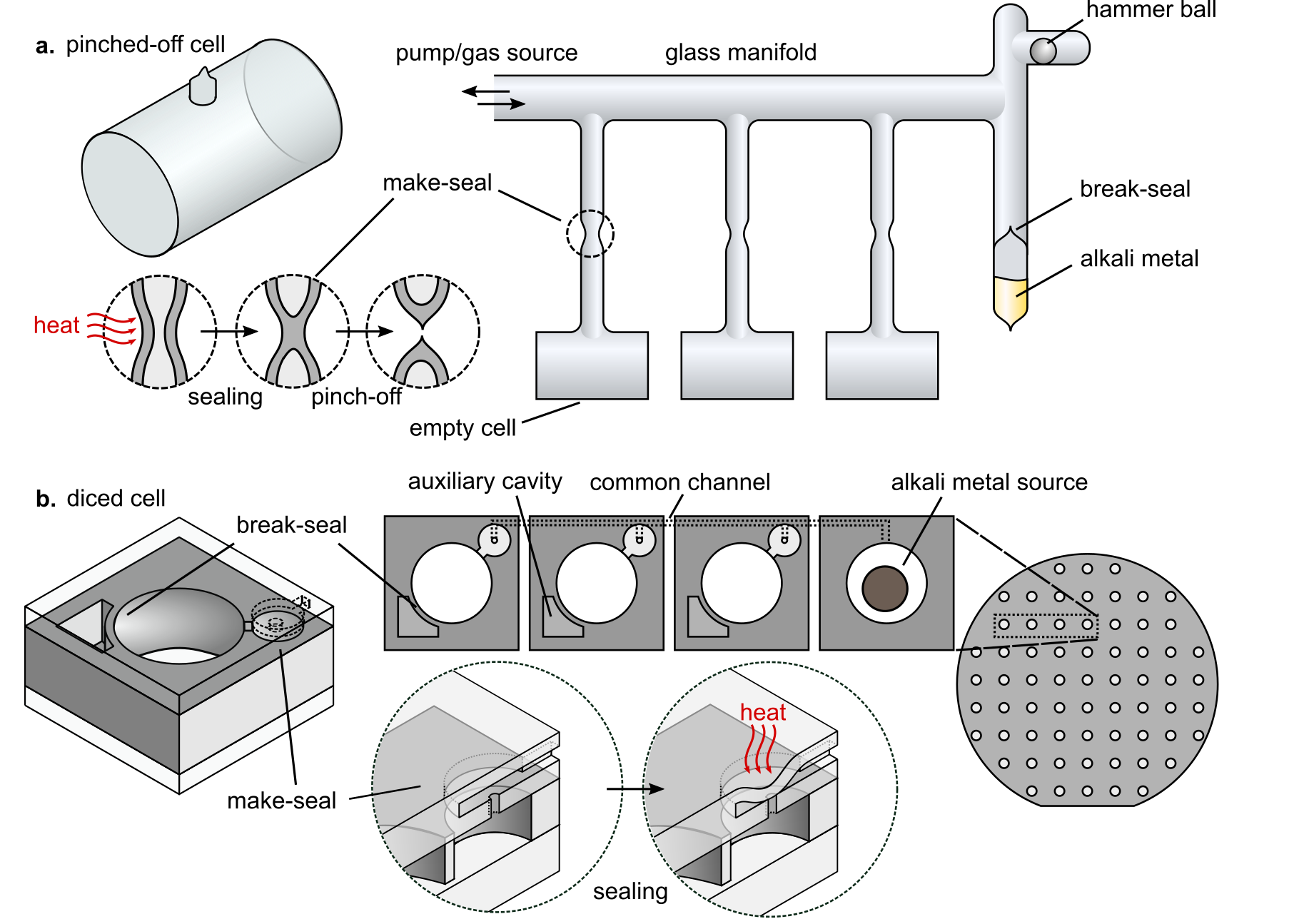}
	\caption{ 
	\textbf{Methods for filling alkali vapor cells.} 
\textbf{a} Conventional alkali vapor cell filling based on glass-blowing.
Empty glass cells are fused to a manifold providing access to an alkali metal source, gas cylinders and a high-vacuum pump.
Sealing is achieved by heating a thin glass capillary, which shrinks due to the pressure difference. Once the channel is fully closed, the cell can be pulled away and pinched-off from the manifold.
\textbf{b} Concept wafer layout integrating laser-actuated make-seal and break-seal structures.
Multiple cells are arranged in an array and connected through a common channel providing access to a single alkali metal source (\emph{e.g.} a dispenser pill).
Cells also feature gas reservoirs initially separated from the main cell chamber.
Integrated laser-actuated make-seal and break-seal structures allow connecting or isolating the different chambers and channel.
After filling and sealing, the individual cells can be released by saw-dicing.
}
	\label{fig:schemes}
\end{figure}

Fig.~\ref{fig:schemes}b shows a conceptual layout implementing make-seal and break-seal structures on a wafer.
This layout provides a way of filling a set of cells from a single alkali metal source and releasing a controlled amount of buffer gas within each cell.
In this layout, multiple cells are patterned in a silicon wafer and are connected through microchannels to the alkali source.
Within each die, a make-seal structure is built so that the main chamber of the cell can be hermetically sealed from the common channel after alkali metal has been introduced by migration from the source.
In addition, auxiliary cavities are patterned around each cell, separated from the main chamber by a break-seal structure.
The auxiliary cavities can be filled to encapsulate a controlled amount of buffer gas at the wafer-level during the first anodic bonding step.
The content of the auxiliary cavity can then be released in the main cavity using the break-seal structure.

We designed make-seal and break-seal structures that can be actuated solely with laser light, once the cell microfabrication is completed, either to melt and fuse the entrance of a channel in the first case, or to ablate a separation wall in the latter case.
Both operations are irreversible but arranging several structures in series or in parallel can circumvent their single-usability.

% MAKE SEAL CONCEPT

Our approach to the make-seal structure consists in deflecting and fusing a thin glass membrane on top of the outlet of a vertical channel.
The structure, shown on Fig.~\ref{fig:schemes}b can be integrated within one of the glass substrates that make up the cells.
To seal the cell, the membrane can be heated up to the point where plastic deformation can occur.
We propose to use a \ce{CO2} laser to heat the membrane locally.
Since both sides of the membrane are subject to different pressures, the membrane is deflected inward until it contacts the channel outlet and fuses with the underlying substrate.
The contact area has an annular shape around the channel, which forms an hermetic bond.

% BREAK SEAL CONCEPT

A break-seal structure can be done by forming a vertical wall isolating two cavities patterned in a silicon substrate, as shown also in Fig.~\ref{fig:schemes}b. 
The wall can then be broken using a laser, allowing a gas to flow across the channel.
Here, we explored a break-seal wall that can be ablated with a pulsed laser without impacting the lid glass and retaining the overall hermeticity of the cells.

Proofs of concept of the two structures are described in the following sections.
We demonstrate their fabrication and provide detailed characterisation for each structure separately, ensuring independent results.

\subsection*{Make-seal}

\begin{figure}[!htp]
	\centering
	\includegraphics[width=15cm]{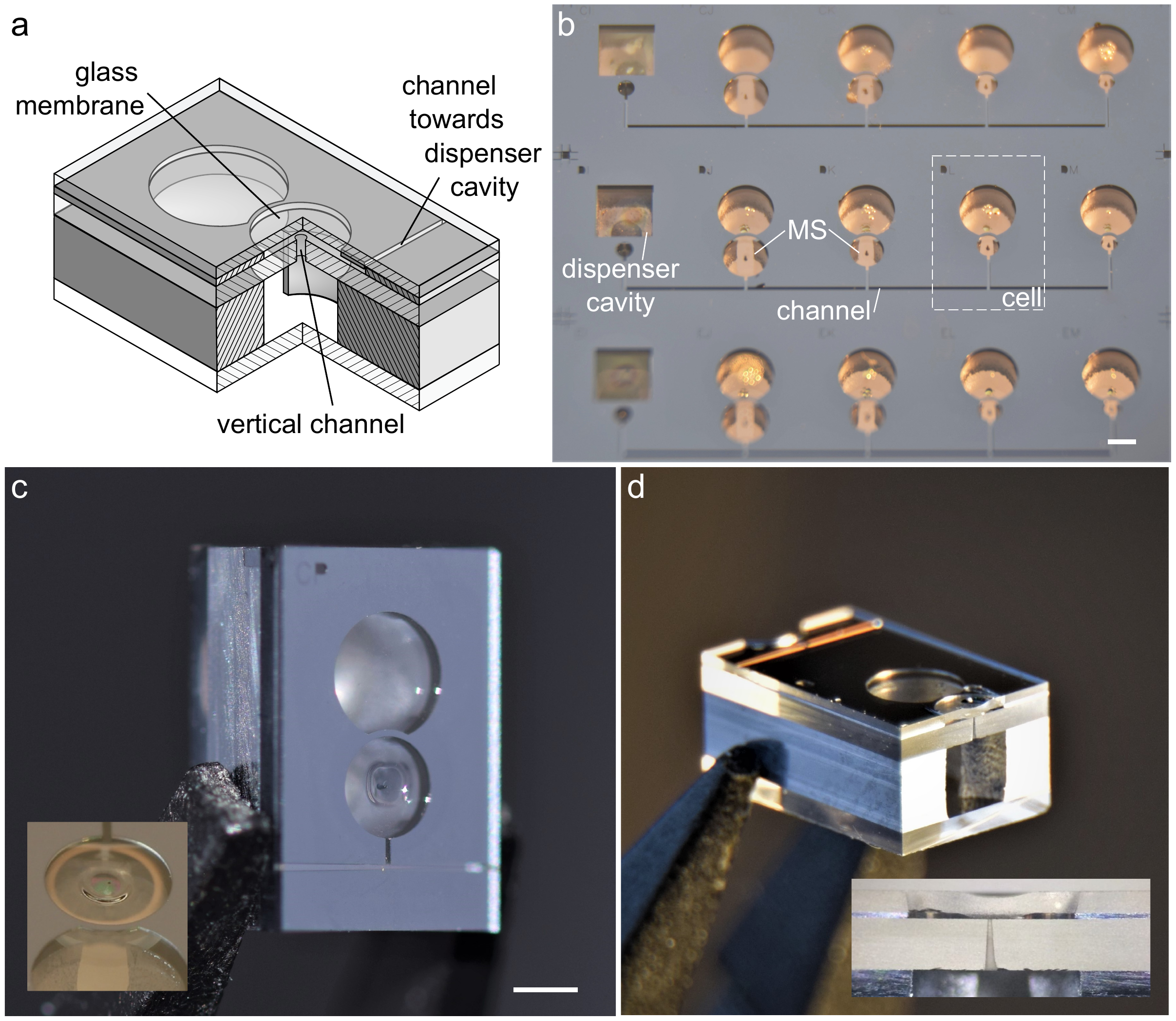}
	\caption{\textbf{Test wafer of make-seal (MS) structures.} \textbf{a} Make-seal demonstrator diagram. \textbf{b} Top-view of a cell wafer with 4-cell clusters having different membrane diameters, all connected to a square dispenser cavity. \textbf{c} Individual cell released by saw-dicing, the channel with \SI{40}{\micro\meter}-diameter mouth is visible under the glass membrane. The latter has been locally heated by laser to deflect it towards the channel mouth and seal it. \textbf{d} Cross-section of a sealed glass membrane obtained by saw-dicing, revealing the sealed vertical channel. The white scale bars are \SI{1}{\milli\meter} long.} 
	\label{fig:MSpics}
\end{figure}

\begin{figure}[!htp]
	\centering
	\includegraphics[width=17cm]{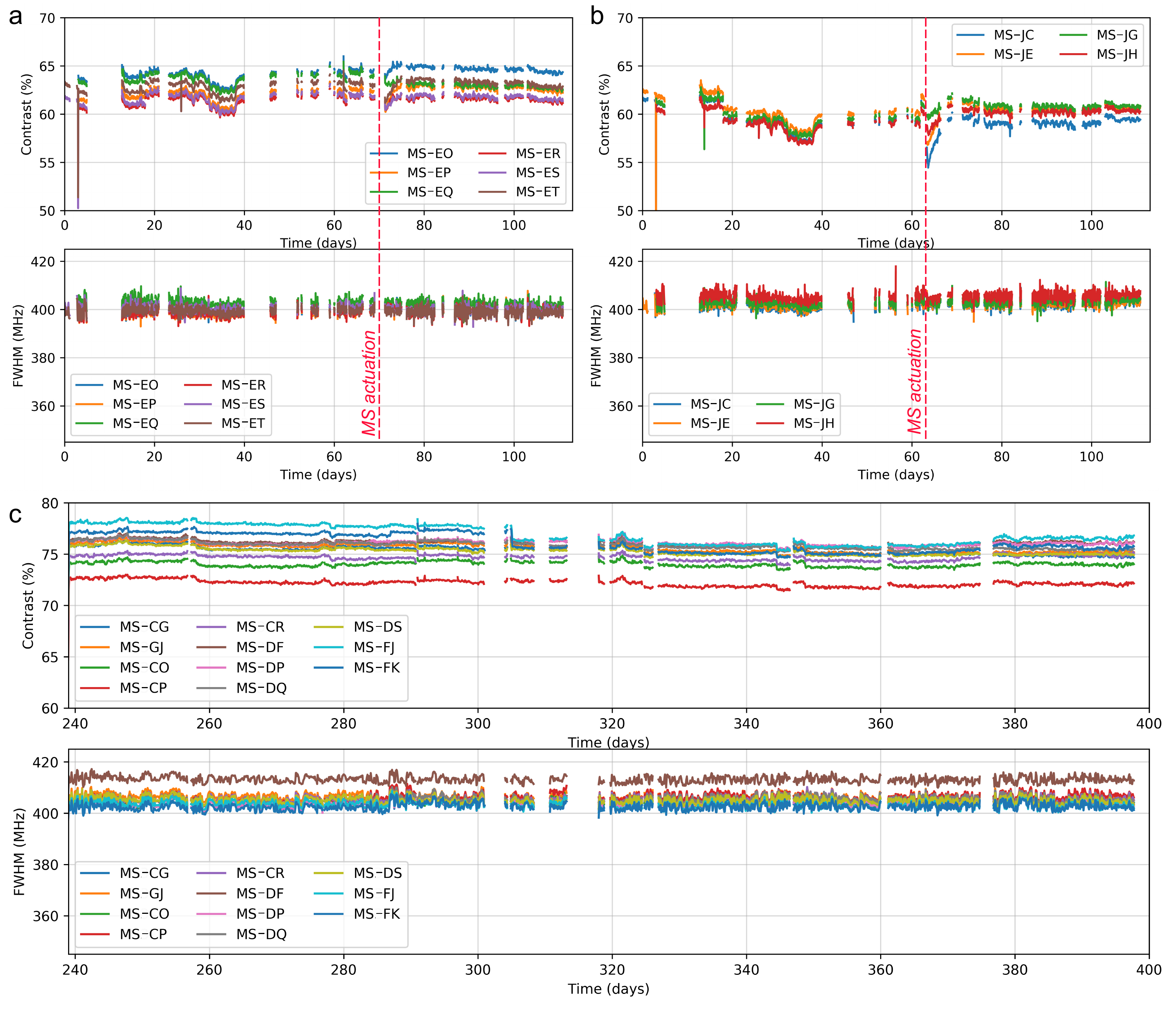}
	\caption{\textbf{Spectroscopic measurements of Cs-vapor cells equipped with make-seal structures.} $D_1$ absorption lines contrast and \ac{FWHM} for  a 6-cell cluster \textbf{(a)} and a 4-cell cluster \textbf{(b)} preceding and following the make-seal actuation.
	Despite a slight and temporary change of the contrast, the values are not affected by laser sealing.
	Note that the measurement gaps correspond to MS actuation campaigns of other cell clusters during which the wafer was removed from the spectroscopy bench.
	\textbf{c}~Evolutions of absorption contrast and \ac{FWHM} during 22~weeks for a dozen of sealed and individually diced cells, including MS-CP shown in Fig.~\ref{fig:MSpics}c.
	Note that the time origin ($t=0~\textrm{days}$) corresponds to the date of activation of the dispensers.
	The graph therefore starts at $t=240~\textrm{days}$ when the cells were diced.
	Average values of contrast and linewidths are slightly larger in \textbf{c} than in \textbf{a} and \textbf{b} due to better thermalization of the individual cells compared to the wafer.} 
	\label{fig:MSgraphs}
\end{figure}

To demonstrate the make-seal structure, we designed a layout in which cells are arranged in clusters.
Each cluster gathers between 3 and 20 cells linked through micro-channels embedded within the cell cap and features a single source cavity hosting a solid alkali metal dispenser (\emph{e.g.} AlkaMax from SAES Getters) (Fig.~\ref{fig:MSpics}a,b).
Channels are built by structuring a silicon wafer used as a spacer between the top cell glass window and an additional glass wafer (Fig.~\ref{fig:MSpics}a,b).
The silicon spacer is through-etched by \ac{DRIE} before bonding and its thickness is adjusted to the desired value by wafer grinding and polishing.
Such structuring allows connecting a cell to its neighbour ones through channels having a section of \SI{40}{\micro\meter}$\times$\SI{100}{\micro\meter} and a \SI{5}{\milli\meter} length.
The path between the channel and the inner cavity of the cell is provided by a \SI{500}{\micro\meter} long vertical channel of \SI{\approx 40}{\micro\meter} diameter and fabricated by femtosecond laser ablation.
Among the test wafer containing about 200 cavities, 20 dispensers were loaded in order to fill 20 clusters made of nearly 100 cells in total.
The final bonding, which seals the cell clusters, is performed under vacuum and the cells are consequently evacuated.
The entire wafer is then loaded on a dedicated optical bench featuring lasers for the activation of the dispensers and for periodic linear absorption measurements.

Once loaded, the dispensers are activated with a high-power laser diode for several seconds, which releases elemental cesium.
The wafer is constantly heated to \SI{70}{\celsius} to let the vapor migrate throughout the cluster.
For each cluster, the cells are laid out in a single line with the source cavity on a far end.
This temperature and this layout let us monitor the filling dynamic for different channel lengths in a non-time-critical manner (Fig.~\ref{fig:MSpics}b).
In such conditions, it typically takes within two to three days before a saturated vapor pressure is established in a cell once its preceding neighbour has been saturated.
Note that the delay to fill a cluster could be shortened with a higher temperature and an optimized arrangement of cells and channels.
As the wafer is heated from the backside, natural air convection tends to yield cold spots at the centre of the main cavity within each cell, where cesium condensation builds up over the course of days (Fig.~\ref{fig:MSpics}b).

Once the cells are filled, they can be hermetically sealed from the common channel and separated from the source cavity thanks to a make-seal apparatus.
The glass membrane is locally heated so that it deflects towards the channel mouth until glass-to-glass fusion bonding and sealing occur.
Fig.~\ref{fig:MSpics}c and d show this deflection.
Local heating is achieved by a \ce{CO2} laser for a few tens of seconds.
The use of such lasers has been proposed in early attempts to make miniature vapor cells, notably by Knappe \emph{et al.} who were able to form glass cells out of thin glass capillaries \cite{knappe_atomic_2003}.
In our case, the laser beam-waist is nearly \SI{200}{\micro\meter} (at $1/e^2$), which limits the extent of the heated area to a portion of the membrane.
Fig.~\ref{fig:MSgraphs}a and b show the evolution of the absorption contrast and linewidth before and after sealing for two different clusters of cells.
No significant variations of contrast or linewidth can be observed.

Our layout includes make-seal membranes of various diameters (Fig.~\ref{fig:MSpics}b), typically ranging from \SI{600}{\micro\meter} to \SI{1600}{\micro\meter} (with a \SI{200}{\um} thickness).
While the large membranes with diameters \SI{1200} and \SI{1600}{\micro\meter} were able to provide long-term sealing (for a third and half of the cells, respectively), smaller ones have been subject to cracks due to excessive stress generated within the reflowed glass.

After sealing the cells, the wafer could be saw-diced into individual cells separated from the source cavity, as shown on Fig.~\ref{fig:MSpics}c.
The resulting cells have a footprint of $\SI{4}{\mm}\times\SI{6}{\mm}$.
A dozen of sealed cells has been monitored for 22 weeks at \SI{70}{\celsius} through linear absorption measurements.
They exhibit a stable atmosphere (Fig.~\ref{fig:MSgraphs}c) with optical lines contrasts and linewidths remaining constant at around \SI{75}{\percent} and \SI{400}{\mega\hertz}, respectively.
The difference in contrast between the measurement performed at the wafer level (Fig.~\ref{fig:MSgraphs}a and b) and at the cell level (Fig.~\ref{fig:MSgraphs}c) is due to differences in temperature, which is not equally accurate for all cell locations throughout the holders.
We can provide a rough upper bound estimate of the leak rate by assuming the effect of the nitrogen intake that would occur in case of a sealing fault. 
Since nitrogen would inflict a broadening of 16~MHz/Torr~\cite{Pitz2009_pressurebroadening}, we can state that no leakage greater than \SI{2}~mTorr/day (\SI{3e-3}{\milli\bar}/day) can be observed with these spectroscopic measurements.

\subsection*{Break-seal}

\begin{figure}[!htp]
	\centering
	\includegraphics[width=15cm]{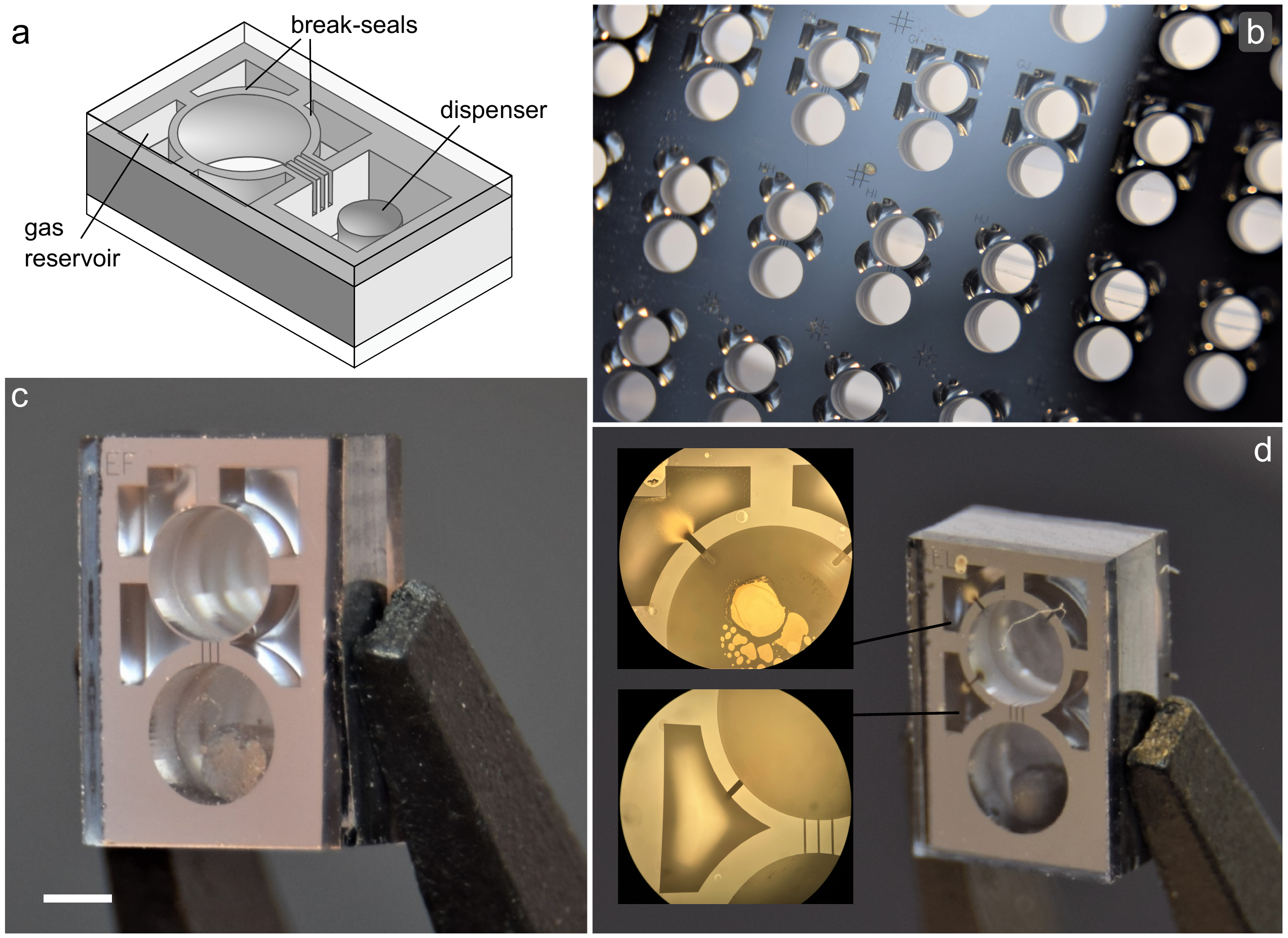}
	\caption{\textbf{Test wafer of break-seal structures.} \textbf{a} break-seal demonstrator diagram. \textbf{b} Top-view of a cell wafer with different reservoir patterns. \textbf{c} Saw-diced individual cell, where the inner walls separating the reservoirs from the main cavity are \SI{50}{\micro\meter} thick. The scale bar is \SI{1}{\milli\meter} long. \textbf{d} Connection between cavities generated through femtosecond laser ablation of inner Si walls. The cell here has been monitored for almost 6 months, during which 3 reservoirs have been sequentially opened as shown in Fig.~\ref{fig:BSgraphs}a.}
	\label{fig:BSpics}
\end{figure}

\begin{figure}[!htp]
	\centering
	\includegraphics[width=15cm]{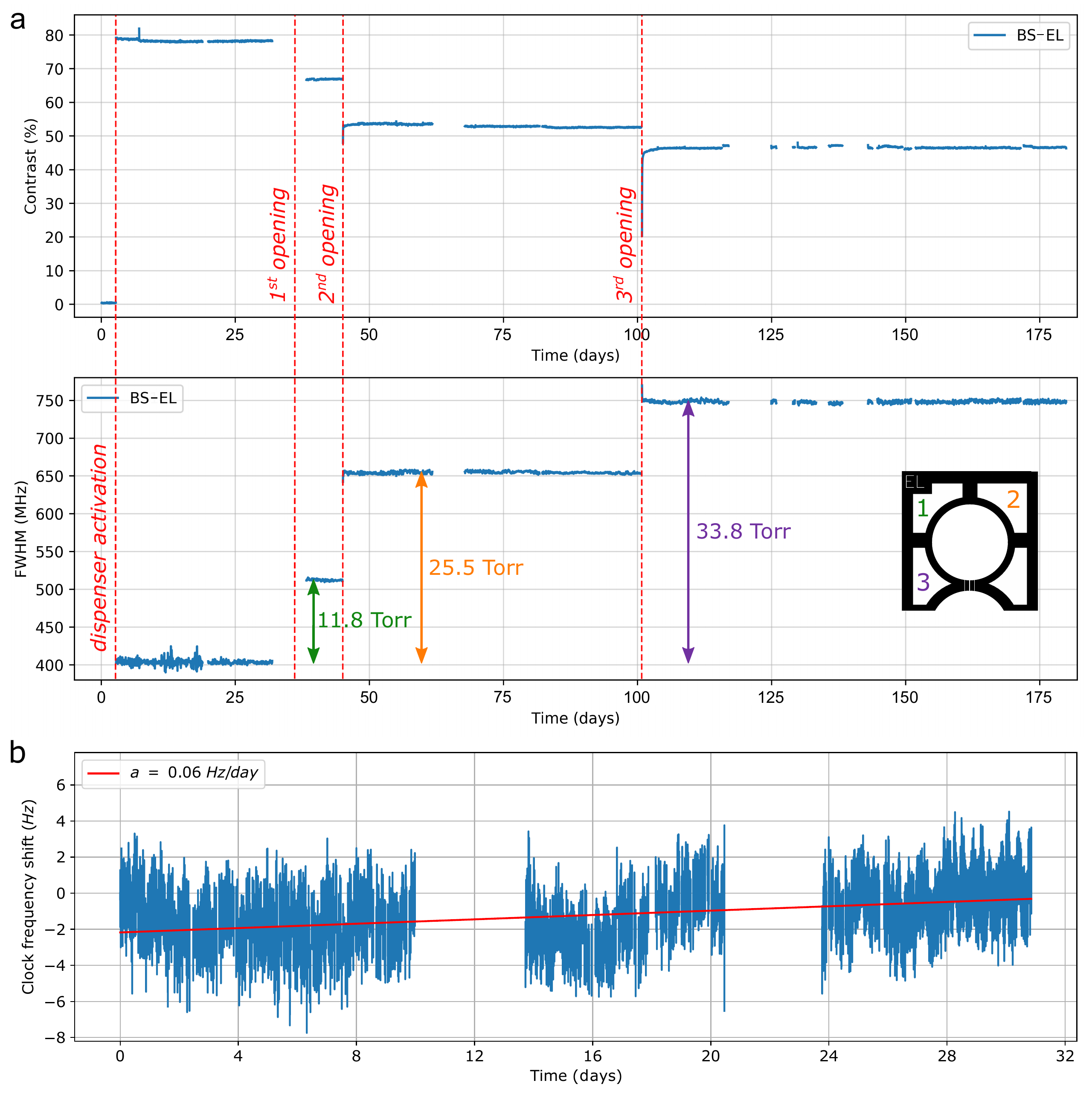}
	\caption{\textbf{Spectroscopic measurements of cesium vapor cells equipped with break-seal structures.} \textbf{a} The cell shown in Fig.~\ref{fig:BSpics}d has been monitored with linear absorption spectroscopy for almost 6 months, during which 3 reservoirs have been sequentially opened. This is well visible in the evolution of contrast and \ac{FWHM} of absorption lines recorded during sequential activation of break-seals, where contrast is decreased and \ac{FWHM} increased stepwise. \textbf{b} The same cell (BS-EL) has afterwards been measured on a \ac{CPT} clock bench for nearly one month. The clock frequency drift is measured at the level of 0.06 Hz/day, \emph{i.e.} 6.5 $\times$ 10$^{-12}$ fractional frequency stability at one day, despite a remaining adjacent gas reservoir filled with 186 Torr of neon.}
	\label{fig:BSgraphs}
\end{figure}

The break-seal structure is demonstrated on cells with the same footprint as the one used for the make-seal, \emph{i.e.} a size of $\SI{4}{\mm}\times\SI{6}{\mm}$.
Here, a collection of non-through etched cavities are arranged around the main cavity to serve as gas reservoirs (Fig.~\ref{fig:BSpics}a).
During the cell assembly, the reservoirs are filled with \SI{186} Torr of neon (pressure derived at \SI{70}{\celsius}).

To fill in the reservoirs, the first anodic bonding between the top glass wafer and the silicon preform is performed under neon atmosphere at \SI{350}{\celsius} with a pressure of several hundreds of torrs so that the target pressure is reached after cool-down.
Subsequently, after introducing the dispenser, the second anodic bonding of the bottom glass wafer is performed under vacuum during which the pressure difference and thus the force applied onto the inner wall can be significant.
The inner walls separating the gas reservoir from the main cavity must therefore be thick enough to handle the pressure difference and provide sufficient hermeticity before breaking, while being thin enough so they can easily be breached.

To explore this parameter space, our wafer layout includes cells with break-seals of different thicknesses, ranging from 50 to \SI{400}{\micro\meter}  (Fig.~\ref{fig:BSpics}b), and 18 cells have been produced from a single wafer.
All of them initially showed a properly evacuated atmosphere within their main cavity, supported by absorption linewidths around \SI{400}{\mega\hertz} at \SI{70}{\celsius}.
We then used a femtosecond ablation laser to open the gas reservoirs sequentially within 5 of these cells (Fig.~\ref{fig:BSpics}c).
Since the laser is focused onto the silicon wall underneath the top glass wafer, the latter remains untouched while the wall is effectively ablated. 
The ablated channels connect the inner cavities of the cells, allowing step-wise pressure adjustments.
Spectroscopic measurements have been performed after each break-seal actuation as shown on the Fig.~\ref{fig:BSgraphs}a for a cell featuring \SI{200}{\micro\meter} thick beak-seals.
First, we activated the dispensers to release pure cesium within the main cavity of the cells.
This increases the contrast to \SI{78}{\percent} and a \SI{405}{\mega\hertz} \ac{FWHM} is measured, matching the values measured in parallel in a large glass-blown evacuated cell kept at room temperature (subtracting the contribution from the temperature difference).
The first break-seal actuation, performed on the upper-left reservoir, releases neon in the main cavities and sets a cell pressure of \SI{11.8}{Torr} (\SI{\pm 0.5}{Torr}).
The next reservoir openings provide step-wise pressure increments to \SI{25.5}{Torr} and \SI{33.8}{Torr}.
Such values, derived from spectroscopic measurements, are in very good agreement with the expected pressures (\SI{12.5}, \SI{25.0} and \SI{33.7}{Torr}) calculated from the amount introduced during fabrication and the ratio of reservoirs and cavities volumes.
This suggests ablation does not cause any major outgassing.
This behavior is observed on the 5 cells under test (although fewer reservoirs have been opened in the other cells).
Consequently, such approach can ensure a control of the inner cell pressure within \SI{1}{Torr}.

In addition to linear absorption, we monitored 3 of the cells in a \ac{CPT} clock configuration to assess the leak rate through a thin untouched break-seal with a greater resolution through the collisional shift of the \ac{CPT} resonance~\cite{kozlova_temperature_2011}.
Among these cells, 2 have been tested after 2 out of 4 reservoirs were opened so that the 2 other reservoirs remained at their initial pressure.

When the cells were measured, the intact reservoirs contained about 186~Torr while the main cavities contained \SI{\approx 20}{Torr}.
For most cells, the \ac{CPT} frequency drift was found to be similar to standard cells (without reservoirs and filled with neon directly), except for the cell featuring a particularly thin break-seal wall (thickness of \SI{50}{\micro\meter}) where a tiny leak could be measured (in the order of \SI{3}~mTorr/day or \SI{4e-3}{\milli\bar}/day).
For a cell featuring \SI{200}{\micro\meter} thick walls (BS-EL), \ac{CPT} clock measurements performed after opening three reservoirs reveal a drift rate on the order of 0.06~Hz/day over 31~days (Fig.~\ref{fig:BSgraphs}b).
This drift rate being small, its uncertainty remains large due to the short and mid-term instabilities of our measurement setups. Yet, the cell behaves similarly to cells having no reservoir so that intact reservoirs could be kept alongside opened ones (in order to tune finely the cell pressure) throughout the cell's lifetime.

%     ____  _                           _           
%    / __ \(_)___________  ____________(_)___  ____ 
%   / / / / / ___/ ___/ / / / ___/ ___/ / __ \/ __ \
%  / /_/ / (__  ) /__/ /_/ (__  |__  ) / /_/ / / / /
% /_____/_/____/\___/\__,_/____/____/_/\____/_/ /_/ 
                                                  
\section*{Discussion}

We have presented a new approach to make, fill and seal microfabricated alkali vapor cells.
It is inspired from conventional cell fabrication techniques based on glass-blowing, which have been used for decades and successfully in many respects.
Featuring microfabricated equivalents to make-seal and break-seal components as key enablers, this approach combines the flexibility of conventional vapor cells and the potential for miniaturisation and mass-production brought forth by well-established MEMS processes.
The make-seal and break-seal devices are actuated by lasers, which can ideally be combined with the laser activation of dispenser to release the alkali vapor, allowing for a complete remote cell-filling process.

We have demonstrated each structure independently and shown how they can be used to fill multiple cells from a single alkali metal source or to backfill the cells with a controlled buffer gas amount. 
Through long-term observations, the hermeticity of both structures was assessed and potential residual leaks found to be below our measurement resolution.
The approach can readily circumvent several of the limitations imposed by current fabrication methods.
First, it allows backfilling cells with nitrogen when solid dispensers are used.
Indeed, nitrogen is commonly used in cell-based atomic clocks, \emph{e.g.} when associated to Ar, to reduce the sensitivity of the clock frequency to temperature variations~\cite{kozlova_temperature_2011} or as a quenching gas to prevent the undesirable effect of radiation trapping~\cite{franz_enhancement_1968}.
Here, nitrogen, or any other gases that would otherwise be absorbed by the dispensers, can be backfilled within the cells.
By allowing on-wafer dispensers or getters to be used and subsequently separated from the cells, applications requiring both high purity and miniature cells such as optical frequency standards could benefit from this method~\cite{newmanArchitecturePhotonicIntegration2019}.
On a research standpoint, we can imagine generating multiple cells with various pressures from a single fabrication run solely by varying the volume of the auxiliary cavities.
This could find uses in pressure-dependent studies and lets us envision pooled fabrication runs or multi-project wafers, common in microelectronics.

Further developments could solve other issues met during integration of atomic devices.
It could indeed be adapted to any atomic or molecular species, but also allow the deposition of temperature-sensitive wall-coatings, or help reaching the high vacuum levels required, for instance, in passively-pumped microfabricated laser-cooling platforms~\cite{boudot_enhanced_2020} or, more broadly, in heat-sensitive \ac{MEMS} devices requiring vacuum encapsulation.

\section*{Materials and methods}

\subsection*{Micro-fabrication}

\begin{figure}[!htp]
	\centering
	\includegraphics[width=15cm]{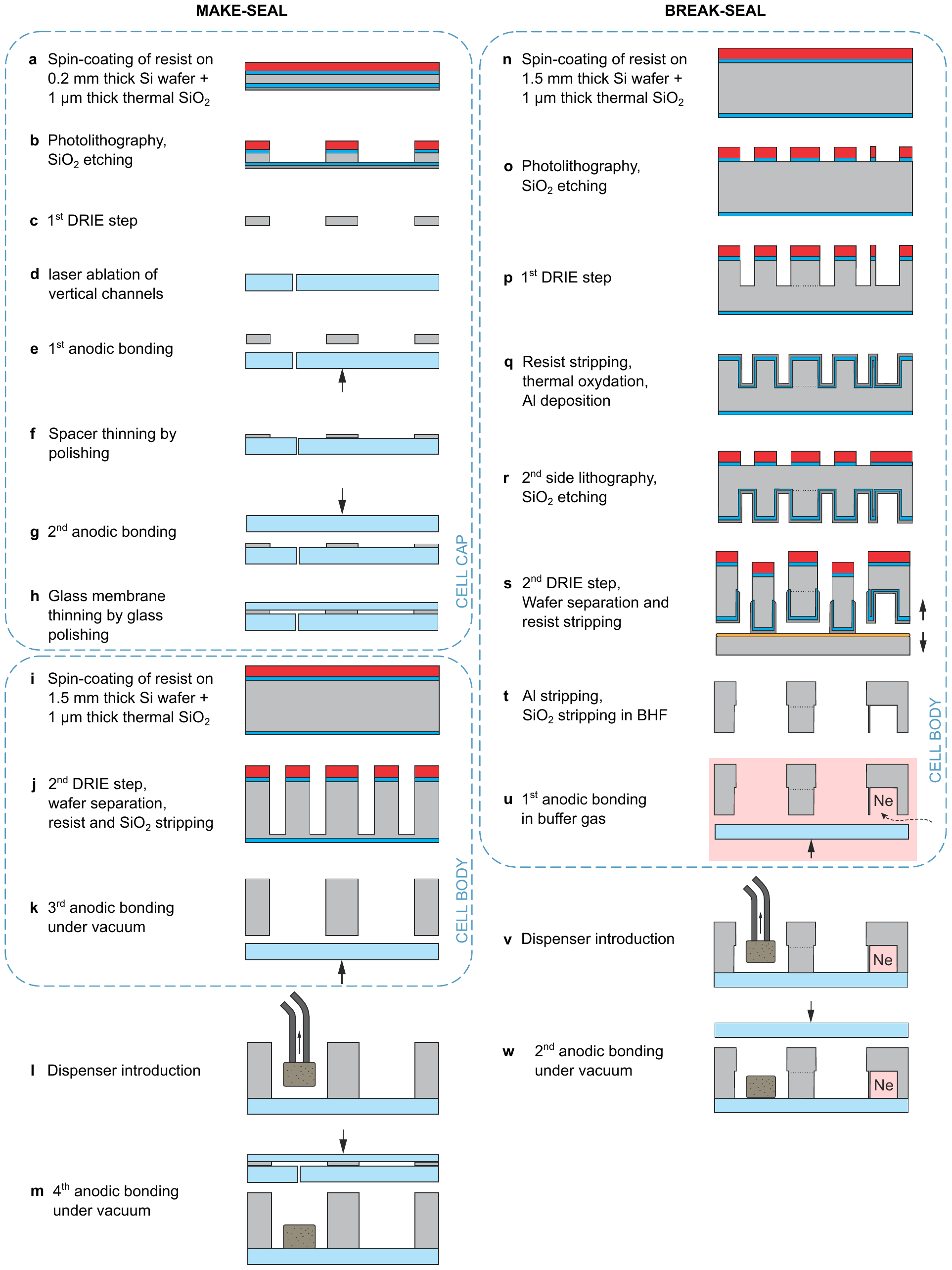}
	\caption{\textbf{Fabrication flow-charts}: Flow-charts of the \textbf{a--m} make-seal and \textbf{n--w} break-seal demonstrators. The processes are applied on 4-inch wafers, which are saw-diced into individual cells once the make-seal is actuated and the cells separated from the dispenser cavity. }
	\label{fig:Flowcharts2}
\end{figure}

The microfabrication flowchart is described in Fig.~\ref{fig:Flowcharts2}.
It relies on well-established processes such as \ac{DRIE} and silicon-glass anodic bonding~\cite{Chutani2014} commonly used to make miniature alkali vapor cells.
On the first hand, the top cell cap that embeds the make-seal structure is made of a glass-Si-glass wafer stack (\SI{100}{\mm} diameter).
To make this cap, the in-plane micro-channels are first etched within a \SI{200}{\micro\meter}-thick Si spacer (Fig.~\ref{fig:Flowcharts2}a-c), which is then bonded to a \SI{500}{\micro\meter}-thick borosilicate glass wafer (Borofloat33 from Schott).
Prior to this bond, the vertical micro-channels are first made in this glass wafer through laser ablation (Fig.~\ref{fig:Flowcharts2}d-e).
After thinning the Si spacer down to \SI{40}{\micro\meter} by lapping and polishing (Fig.~\ref{fig:Flowcharts2}f), a second glass wafer is bonded and in turn thinned down to \SI{200}{\micro\meter} (Fig.~\ref{fig:Flowcharts2}g). The cap with the interconnecting channels and glass membranes is thereby created (Fig.~\ref{fig:Flowcharts2}h).
The cell body is formed by a through-etched \SI{1500}{\micro\meter}-thick Si wafer where cavities are etched and bonded to another  \SI{500}{\micro\meter}-thick glass wafer (Fig.~\ref{fig:Flowcharts2}i-k). Once the dispensers are loaded, the final anodic bonding is performed under vacuum (Fig.~\ref{fig:Flowcharts2}m).
Although the proofs of concept of the two different seals are presented in here separately, mostly for complete and independent characterization (of leaks and long-term atmosphere stability), the cell body could easily be fabricated following the break-seal flow-chart.
In such a case, the \SI{1500}{\micro\meter}-thick Si wafer is not through-etched from one side, instead, it is processed with double-sided lithography and etching in order to etch-through only the main cavities whereas the reservoir cavities are nearly \SI{800}{\micro\meter} deep (Fig.~\ref{fig:Flowcharts2}n-t).
This approach allows trapping gas during the first anodic bonding (former third bonding in the make-seal flow-chart) (Fig.~\ref{fig:Flowcharts2}u). Then, the final anodic bonding can be performed under vacuum (Fig.~\ref{fig:Flowcharts2}w).
It can be noticed that it could also be performed in a dispenser-friendly gas environment while the preceding bonding would trap an incompatible gas, eventually released once the make-seal has been actuated in order to reach a well-controlled gas mixture.

\subsection*{Characterization benches}

The cells are primarily characterized by a linear absorption spectroscopy setup.
The bench is equipped with several lasers: a high power laser diode ($\lambda=\SI{808}{\nano\meter}$) for dispenser activation, and a distributed feedback laser resonant with the $D_1$ lines of Cs.
The cells can be mounted on heating holders (usually maintained at \SI{70}{\celsius}) able to carry a wafer and nearly a hundred of individual cells, which are moved by an XY-stage.
This stage allows sequential measurements of each cell at regular intervals (here, spectra are recorded every 15 minutes).
For each measurement, the absorption signal is normalized by the sloped background resulting from the ramp applied to the laser injection current. Hence, its logarithm is fitted with double Gaussian or double Voigt profiles, depending on the presence of a buffer gas.
The contrast and \ac{FWHM} of the absorption lines can thus be derived.
Such features allow monitoring the atomic density as well as the buffer gas pressure.
As it can be seen on the Fig.~\ref{fig:BSgraphs}e, the contrast and the \ac{FWHM} is typically measured within \SI{\pm 0.4}{\percent} RMS and \SI{\pm 4}{\mega\hertz}, respectively.
Note that the uncertainty is twice as low for individual cells than for wafer-level measurements, thanks to a better temperature control during the spectroscopic measurements.
For measurements lasting several months and considering \emph{e.g.} the coefficient of collisional broadening for neon of \SI{10.36}{\mega\hertz}/Torr~\cite{Pitz2009_pressurebroadening} at \SI{70}{\celsius}, it is then possible to detect gas leaks in the order of several mTorr/day, provided that the measurements are long enough.

For smaller leaks, the linear absorption spectroscopy setup is not accurate enough and we consequently rely on a \ac{CPT} clock bench.
Within several days, we can reasonably detect frequency shifts below 0.5~Hz/day, which corresponds to \SI{\approx 1}~mTorr/day of buffer gas pressure variation (neon collisional shift of the clock frequency of 686~Hz/Torr~\cite{kozlova_temperature_2011}). The setup for more resolved measurements is described in more details by Vicarini \emph{et al.}~\cite{Vicarini2018}.

Two laser benches are involved for actuating the seals, relying either on a CO$_2$ laser ($\lambda=\SI{10.6}{\micro\meter}$) or a femtosecond laser ($\lambda=\SI{1030}{\nano\meter}$).
Actuation of both make-seals and break-seals requires several seconds to several tens of seconds per cell.

\bibliography{Maurice_sample}

\section*{Acknowledgements}
This work was supported partially by the D\'{e}l\'{e}gation G\'{e}n\'{e}rale de
l’Armement (DGA) and by the Agence Nationale de la Recherche (ANR) in the frame of the ASTRID project named PULSACION under
Grant ANR-19-ASTR-0013-01. It was also supported by the EIPHI Graduate school (Grant ANR-17-EURE-0002) and by the R\'{e}gion Franche-Comt\'{e} (NOUGECELL project). The work of Cl\'{e}ment Carl\'{e} was supported by the Centre National des Etudes Spatiales (CNES) and the Agence Innovation D\'{e}fense (AID).
The authors thank the French RENATECH network and its FEMTO-ST technological facility, in particular Sylwester Bargiel for his assistance with anodic bonding.
They also want to thank Femto-Engineering for the support with laser ablation.
They are finally grateful for Pierre Bonnay and St\'{e}phane Gu\'{e}randel from SYRTE for introducing the glass-blown methodology as well as Olivier Gaiffe for fruitful discussions.

\section*{Conflict of interests}

The authors declare no competing interests.

\section*{Contributions}

V.M. and N.P. initiated and conceived the proof-of-concept experiments, V.M., R.C. L. G.-M. designed and fabricated the make-seal devices, S. Q. and S.K. and N. P. designed and fabricated the break-seal devices, V.M., J.-M. C., R. V. and N. P. built the optical benches, C.C., M. A. H., R. B. and N.P. characterized the devices, V. M., C. C., R. B . and N. P. analyzed the results, V. M. and N. P. wrote the manuscript. All authors reviewed the manuscript.

\end{document}